# TOLERANCING ANALYSIS AND FUNCTIONAL REQUIREMENT

**Petit Jean-Philippe**
Laboratoire de Mécanique Appliquée (LMécA / ESIA), BP 806 74016 ANNECY Cedex,
+33 04 50 09 65 71, +33 04 50 09 66 49, jean-philippe.petit@esia.univ-savoie.fr
**Serge Samper**
Laboratoire de Mécanique Appliquée (LMécA / ESIA), BP 806 74016 ANNECY Cedex,
+33 04 50 09 65 64, +33 04 50 09 66 49, serge.samper@esia.univ-savoie.fr

**Abstract:**

*The aim of this paper is to show through a simple assembly a method of tolerancing analysis (coherent with GPS) developed at LMécA and based on the model of clearance and deviation domains. Tolerancing is an important step in the product design because on it will depend the functionality of the mechanism its assemblibility but also its cost: manufacturing cost increases with the precision of tolerances values. In our model, each feature specification is translated into tolerance zone. This zone limits displacements of the toleranced feature. Limitations of this small displacements are expressed in a mathematical form by a 6-polytope in a 6 dimensions space (3 rotations and 3 translations). In the same way, contact conditions in joints allow to write linear inequalities which can be translated by 6-polytopes. Each domain is defined by a set of vertices and a system of inequalities. Considering a chosen tolerancing, the method allows to verify the mechanism assemblibility but also several functional requirements. The example of a minimal clearance requirement between two surfaces will be treated. We will show the residual clearance zone associated to these surfaces considering tolerances on parts and clearances in joints. With such a tool, the designer will be able to modify values of the tolerances and thus to reduce the manufacturing cost while guaranteeing the functionality of the mechanism.*

## 1 Introduction

A mechanism is composed of manufactured parts which are imperfect. To answer to functional requirements, the designer has to define each part dimensions and the limits of its geometry variations. This work is decomposed into two steps:
- Dimensioning (definition of the perfect nominal geometry).
- Tolerancing (definition of the authorized deviations with regard to the nominal geometry).

Tolerancing analysis means validating designer choices, so as to verify that each tolerance expression respects the standard and the functional requirements (assemblability, precision, non interference…). Different works concern this aspect of analysis, involving various methods: several authors [1] [2] use 3D polytopes to represent a tolerancing. From our point of view, it is necessary to consider specifications chosen by the designer but it is also indispensable to take into account clearances into joints. These considerations lead on the construction and the use of polytopes [3] in a 6-dimension space (cuttings of these hypervolumes allow a 3D display).

In this paper, an analysis method [4] based on a representation tool of functional specifications and joints clearances will be discussed.

## 2  Model and hypothesis

The model, called model of the clearances and deviations domains, developed in the LMécA Research Laboratory is based on various hypotheses:
o  Parts constituting the mechanism are supposed rigid.
o  Form defects are not considered: real features are modelled by elementary features (plan, cylinder, cone…) called associated features (Figure 1). Various criteria allow this replacement, for example the Gauss criterion.
o  Other types of defects are considered: dimension defects (relative to dimension value, thus to dimensioning) are parameters to be considered during the analysis. Position and orientation defects are considered as displacements between the frames attached to associated features and nominal features. Assuming that these displacements are small enough, the position and the orientation of a frame with regard to another one are translated by a small displacement torsor [5] called the deviation torsor.

The general form of a deviation torsor is written: $\{E\} = \begin{Bmatrix} Tx & Rx \\ Ty & Ry \\ Tz & Rz \end{Bmatrix}$  (1)

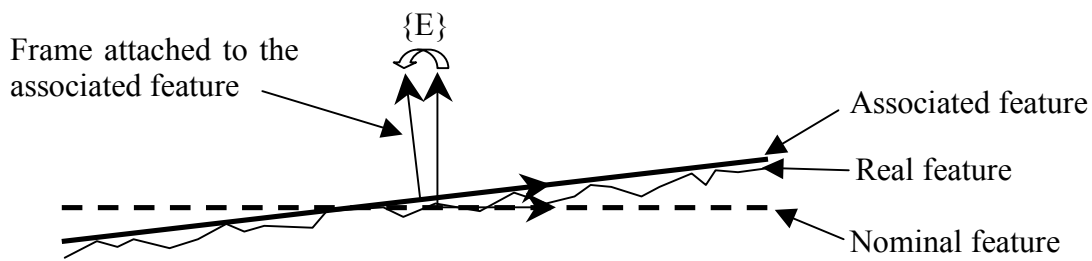

*Figure 1.  Real feature and associated feature.*

### 2.1 Deviation domain

Standards allow to represent each geometric specification by a tolerance zone (Figure 2). This zone limits toleranced feature displacements with regard to one or several references. This means imposing the limits of the deviation torsor between the frame attached to the associated feature and the frame built on the reference system.

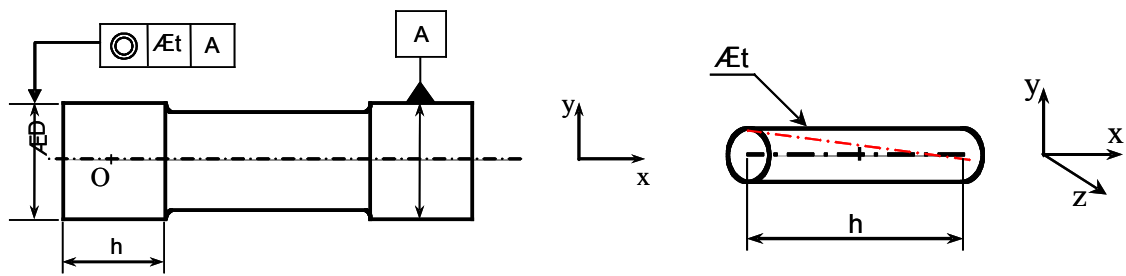

*Figure 2. Specification example with corresponding tolerance zone.*

So the deviation domain limits characterize small displacements which are attached to this tolerance. In general, these limits are translated by a system of linear inequalities (eq. 2) which concern 6 components (3 rotations and 3 translations).

When inequalities are not linear, it is possible to make a linear approximation. Then, this system is translated in a 6-dimension domain in the shape of a hypervolume called the deviation domain [6] and noted *[E]*. A double definition of this 6-polytope is used: a set of linear inequalities and a list of the domain vertices.

$$\begin{cases} \dfrac{t}{2} - Tz - \dfrac{h}{2} \cdot Rz \geq 0 \\ \dfrac{t}{2} - Tz + \dfrac{h}{2} \cdot Rz \geq 0 \\ \ldots \\ \dfrac{t}{2} + 0 \cdot Tx + \dfrac{\sqrt{3}}{2} \cdot Ty + \dfrac{1}{2} \cdot Tz + 0 \cdot Rx - \dfrac{1}{4} \cdot h \cdot Ry + \dfrac{\sqrt{3}}{4} \cdot h \cdot Rz \geq 0 \\ \dfrac{t}{2} + 0 \cdot Tx + \dfrac{\sqrt{3}}{2} \cdot Ty + \dfrac{1}{2} \cdot Tz + 0 \cdot Rx + \dfrac{1}{4} \cdot h \cdot Ry - \dfrac{\sqrt{3}}{4} \cdot h \cdot Rz \geq 0 \end{cases} \quad (2)$$

The polytope below (Figure 3) is a 3D cut of the coaxiality resulting domain shown in Figure 2 and expressed at the point O (with *h = 10* and *t = 0.05*). This 3D representation can be done by cutting the 6-polytope through 3 directions. 3 components are fixed here: *Tx = Ty = Rx = 0*. Each domain is defined by a minimal set of inequalities (eq. 2) and a list of its vertices whose coordinates characterize maximum displacements of the toleranced feature inside the tolerance zone.

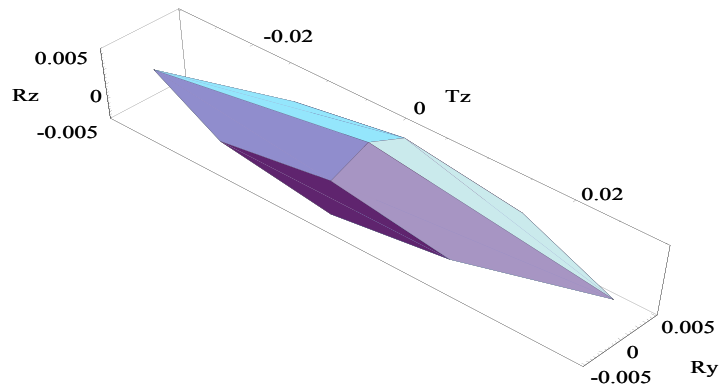

*Figure 3. Deviation domain.*

## 2.2 Clearance domain

Une liaison est constituée de deux pièces. Ces pièces n'étant pas parfaites, il est nécessaire de faire apparaître du jeu entre celles-ci pour rattraper les défauts de fabrication. De la même façon qu'un domaine écart caractérise les déplacements admissibles de l'élément tolérancé dans la zone de tolérance qui lui est associée, un domaine jeu (noté [J]) définit l'ensemble des petits déplacements permis par une liaison. Pour cela, un repère est construit sur chacune des pièces et les déplacements d'un repère par rapport à l'autre autorisés par les conditions de contact dans la liaison permettent d'écrire un système d'inéquations linéaire ainsi que la liste des coordonnées des sommets définissant ce domaine dans un espace de dimension 6.

**Remarque** : un domaine jeu est infini dans les directions correspondantes aux degrés de liberté de la liaison. Pour une liaison pivot glissant d'axe $\vec{x}$ par exemple, le domaine jeu résultant sera borné suivant Ty, Tz, Ry et Rz mais sera infini suivant Tx et Rx.

## 2.3 Geometric operations on domains

En fonction de la configuration du mécanisme (boucles simples, parallèles, ouvertes…) et en s'appuyant sur la théorie des mécanismes, la méthode d'analyse de tolérancement fait appel à différentes opérations géométriques sur les domaines jeux et écarts donc sur des 6-polytopes. Ces opérations sont : la somme de Minkowski exprimée par le symbole ⊕ (voir Figure 4), l'intersection de domaines ainsi que la vérification d'inclusion d'un domaine à l'intérieur d'un autre.

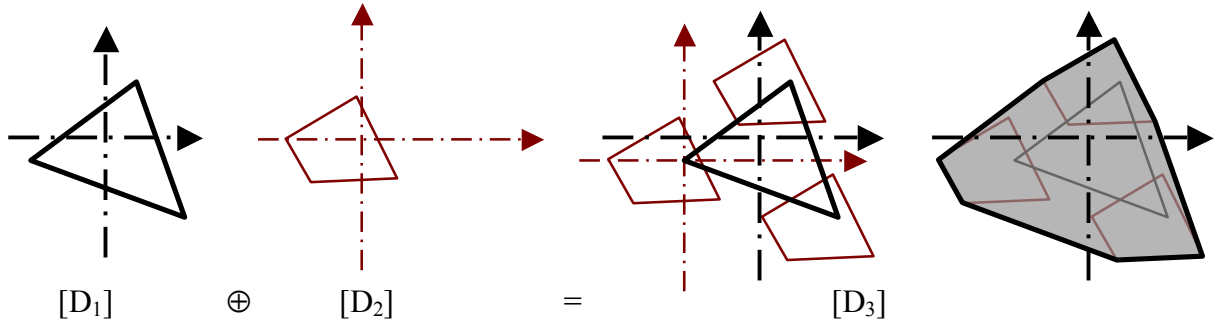

*Figure 4. Minkowski addition of two convex domains.*

## 3 Residual clearance domain

Lors de l'assemblage de deux pièces, les surfaces de contact sont dites fonctionnelles et le concepteur doit donc définir les écarts de géométrie maximums sur ces surfaces à travers ses choix de tolérancement. *A* étant la surface de liaison entre les pièces *0* et *1* (Figure 5) il est possible de calculer le torseur écart $\{E_{0A}\}$ entre le repère $F_0$ attaché à la pièce *0* et le repère $F_{0A}$ construit sur $A_0$.

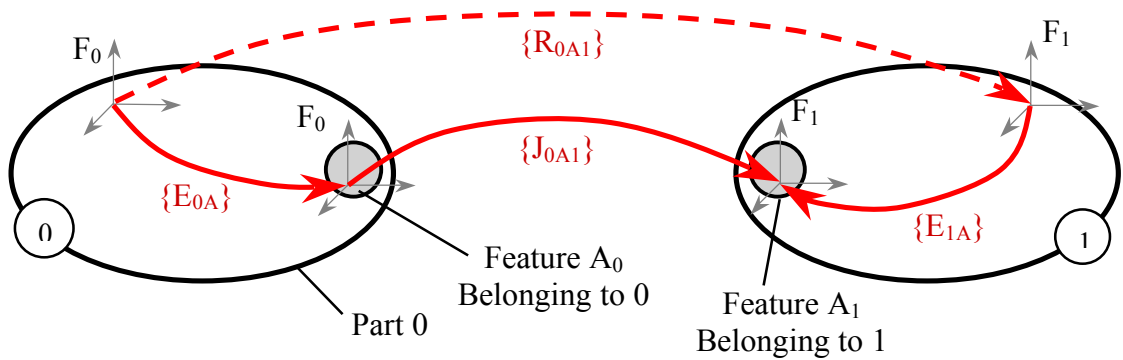

*Figure 5. Diagram of the assembly of two parts.*

L'ensemble des valeurs de ce torseur définit les déplacements de la surface tolérancée dans sa zone de tolérance, il peut être exprimé sous la forme du domaine écart $[E_{0A}]$. $\{J_{0A1}\}$ est le torseur jeu relatif au contact au sein de la liaison entre les pièces *0* et *1*. Considérant les défauts sur les surfaces usinées mais limités par le tolérancement, il est possible de déterminer le domaine jeu résiduel $[R_{0A1}]$ de la liaison de la manière suivante :

$$[R_{0A1}] = [J_{0A1}] \ominus [[E_{0A}] \oplus [E_{A1}]] \tag{3}$$

**Remarques** :

- Contrairement à la somme de Minkowski, l'opération $\ominus$ n'est pas commutative.
- Si les pièces sont parfaites, le jeu résiduel est égal au jeu nominal de la liaison.
- Si le domaine jeu résiduel d'une liaison existe alors l'assemblage des pièces constituant cette liaison sera toujours possible. Ceci provient du fait que les défauts cumulés sur les surfaces en contact sont pris en compte (on se place dans le pire des cas).
- Géométriquement, cette opération consiste à garder l'intersection des domaines jeux « balayés » sur la somme des écarts (Figure 6).
- Le domaine jeu résiduel représente le jeu minimum garanti pour la liaison et ce quels que soient les défauts sur les surfaces usinées (mais limités par le tolérancement).
- Le domaine $[J_{0A1}] \oplus [[E_{0A}] \oplus [E_{A1}]]$ indique la précision de l'assemblage. Les bornes de ce domaine indiquent les jeux maximums que pourrait présenter la liaison constituée de deux pièces quelconques appartenant à un lot.

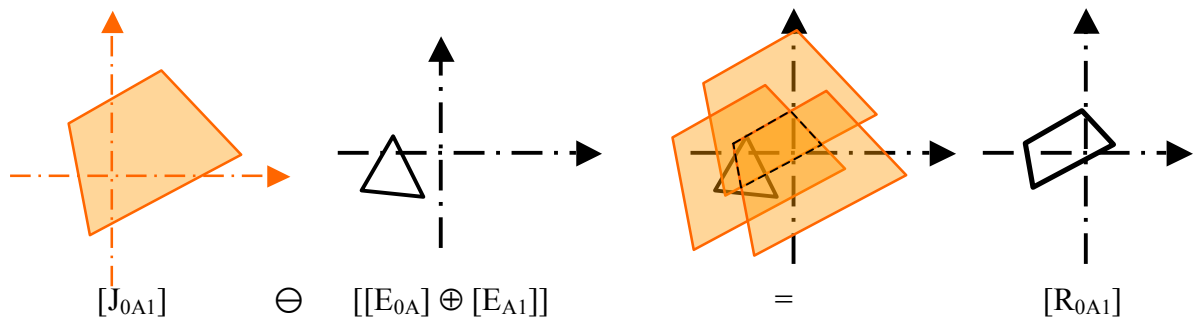

$[J_{0A1}] \quad \ominus \quad [[E_{0A}] \oplus [E_{A1}]] \quad = \quad [R_{0A1}]$

*Figure 6. Construction of a residual clearance domain*

## 3.1 Example

## 4 Conclusion

## References

[1] D. TEISSANDIER, V. DELOS, Y. COUETARD. *"Operations on polytopes: application to tolerance analysis"*, Proceedings of the 6th CIRP International Seminar on Computer-Aided Tolerancing, Fred van Housten/Hubert Kals, 1999, pp. 425-434.

[2] S. BHIDE, J. K. DAVIDSON, J. J. SHAHA. *"Areal Coordinates: The Basis of a Mathematical Model for Geometric Tolerances"*, Proceedings of the 7th CIRP International Seminar on Computer-Aided Tolerancing, P. Bourdet L Mathieu, 2001, pp. 83-92.

[3] K. FUKUDA. *cdd, cddplus and cddlib homepage.* McGill University, Montreal, Canada, 2002. http://www.cs.mcgill.ca/~fukuda/software/cdd home/cdd.html.

[4] M. GIORDANO, E. PAIREL, S. SAMPER. *"Mathematical representation of Tolerance Zones"*, Proceedings of the 6th CIRP International Seminar on Computer-Aided Tolerancing, Fred van Housten/Hubert Kals, 1999, pp. 177-186.

[5] D. GAUNET. *"Modèle Formel de Tolérancement de Position. Contribution à l'Aide au Tolérancement des Mécanismes en CFAO"*, PHD Thesis at Laboratoire de Mécatronique de l'ISMCM, 1994, pp. 23-25.

[6] M. GIORDANO, D DURET. *"Clearance space and deviation space. Application to three dimensional chain of dimensions and positions"*, 3rd CIRP International Seminar on Computer-Aided Tolerancing, 1993, pp. 179-196.